# Hypersoft X-ray Sources as a Low-Energy Class of Luminous Cosmic Emitters

Mustafa Muhibullah[1*], Jimmy A. Irwin[1†], Rosanne Di Stefano[2†]

[1] Department of Physics and Astronomy, University of Alabama, Box 870324, Tuscaloosa, 35487, AL, USA.

[2] Center for Astrophysics, Harvard & Smithsonian, 60 Garden St., Cambridge, 02138, MA, USA.

* Corresponding author. Email: mmuhibullah@crimson.ua.edu;

† These authors contributed equally to this work.

**X-ray binaries, powered by accretion onto black holes, neutron stars, or white dwarfs from companion stars, are among the brightest non-explosive beacons in galaxies, outshining the Sun by millions. Typical X-ray binaries emit primarily above 0.3 keV, whereas cooler sources, peaking in the extreme ultraviolet (EUV), remain challenging to detect due to a long-standing observational blind spot. Here, we report the discovery of a remarkable class of luminous, point-like, non-nuclear X-ray objects—hypersoft X-ray sources—that have largely been missed by previous surveys. Detected primarily or exclusively below 0.3 keV, the most luminous examples radiate near $10^{38}$ erg/s in the narrow X-ray band, with spectral models suggesting even higher bolometric luminosities, most of which emerge in the EUV. They rank among the most energetic sources in galaxies, yet their EUV-peaking spectra evaded earlier detections. We propose that hypersoft sources represent X-ray binaries spanning multiple physical classes, including accreting white dwarfs or post-nova systems—potential progenitors of Type Ia supernovae—and systems hosting accreting black holes. Beyond their elusive nature, they may play a crucial role in ionizing gas within galaxies.**

The extreme ultraviolet (EUV) window, spanning 100–912 Å (13.6 eV–0.124 keV), occupies a critical yet largely uncharted territory of the electromagnetic spectrum[1]. Despite its importance for understanding hot astrophysical sources, the neutral hydrogen and helium in the interstellar medium absorb EUV photons so efficiently that they form a nearly impenetrable barrier[2,3]. While the EUV band remains largely inaccessible, EUV-peaking spectra would be expected to emit a detectable soft X-ray tail of emission that offers a potential indirect window into this hidden domain. Even so, detecting EUV-peaking sources via their soft X-ray tail is inherently challenging: lower-energy X-ray photons are also susceptible to extinction by Galactic gas, and most current X-ray detectors exhibit rapidly declining sensitivity below 0.5 keV.

We conducted a systematic search for extremely soft X-ray sources that may represent the high-energy tail of a hidden EUV source population. We searched in six well-observed, nearby external galaxies using archival *Chandra* data. These galaxies were selected for their long exposure times, proximity, and diversity in galaxy type, optimizing the chances of uncovering these exceptionally soft X-ray emitters. By imposing stringent selection criteria, we uncovered a previously unrecognized population of sources that emit primarily or exclusively below 0.3 keV. We refer to these objects as hypersoft X-ray sources (HSSs)—a newly revealed class that radiates at the lowest energies accessible to modern X-ray instrumentation, making them among the softest X-ray sources yet observed. To identify HSSs, we required $\geq 3\sigma$ detections in the 0.15–0.3 keV energy range, with no significant emission above 0.3 keV. To ensure a more conservative classification, we impose an additional criterion that the photon ratio between the 0.15–0.3 keV and 0.3–1.0 keV bands must exceed ~8 (see *Methods*). All candidate sources were visually inspected to exclude spurious detections arising from elevated background levels and/or background flares.

For two galaxies, NGC 3379 and NGC 4472, Fig. 1 illustrates that the HSSs are bright below 0.3 keV but exhibit no emission above that value. This name is given to contrast them to the somewhat harder supersoft X-ray sources (SSSs)[4], which emit energies up to around 1.0 keV[5,6]. Like HSSs, SSSs are phenomenologically defined by their ranges of temperatures and luminosities. The class includes black hole (BH) and white dwarf (WD) accretors, either actively accreting material from a companion star[7], or in a post-nova phase with residual hydrogen shell burning[8], and potentially neutron stars.

We discovered 84 HSSs, with 7 to 21 in each galaxy (Table 1), found in both star-forming and older galactic populations. While a few, nearby low luminosity HSSs were previously known X-ray sources, nearly all of the more luminous and distant HSSs remained undiscovered, despite decades of X-ray observations. This is because the current generation of X-ray telescopes either lack the spatial resolution or the collecting area below 0.3 keV to have readily detected these sources previously. Detectability of HSSs depends strongly on both source location and instrumental sensitivity; as a result, the number of detected HSSs likely represents a lower limit of the true underlying population within each host galaxy. A comprehensive list of source properties is provided in Extended Data Table 1.

We estimate that HSSs have unabsorbed 0.15–0.3 keV luminosities ranging from $10^{35}$-$10^{38}$ erg $s^{-1}$, assuming isotropic emission, depending on the detection limits of each galaxy (see *Methods* for spectral fitting details). The lowest-luminosity HSSs are detectable only in our nearest galactic neighbor, M31. Sources with $10^{36}$-$10^{37}$ erg $s^{-1}$ can be found in M101, which is closer than the other galaxies on our list, and which has had a very deep (~1 megasecond) exposure. These are the only spiral galaxies we have studied. Given the greater distances and relatively shorter exposures of the other galaxies on our list, which are early-type galaxies, only the brightest HSSs can be detected.

If the emission is anisotropic or beamed, the isotropic-equivalent luminosities reported here would overestimate the intrinsic luminosities. Strong beaming would also make such systems observable only over a restricted range of viewing angles, potentially implying a larger underlying population than is directly detected. As the spectral energy distributions (SEDs) of these sources likely peak in the EUV (Extended Data Fig. 2), their total bolometric energy output must be markedly higher than these limited bandwidth values.

Although precise temperature measurements are not possible due to low count rates and calibration uncertainties below 0.3 keV in *Chandra*, the absence of significant emission in the harder 0.3–1.0 keV band strongly indicates that these sources have exceptionally low effective temperatures, likely not exceeding ~24 eV (280,000 K). This upper limit does not rely on direct spectral fitting of the HSSs, which is unreliable in the 0.15–0.3 keV range, but is instead motivated by the well-studied Milky Way supersoft source RX J0439.8−6809. Despite its spectroscopically measured temperature of only 21.5 ± 2.6 eV (250,000 ± 30,000 K)[9], RX J0439.8−6809 exhibits detectable emission in the 0.3–1.0 keV band (see *Methods*), whereas the HSSs do not. Consistent with this, a lower-luminosity source in M31 previously observed with *XMM-Newton* has blackbody temperature of $17^{+7}_{-6}$ eV ($200{,}000^{+80{,}000}_{-70{,}000}$ K; M31N 1996-08b)[10,11], representing the only HSS with a published temperature estimate to date.

Theoretically, a ~24 eV blackbody spectrum implies a bolometric correction factor of ~8 to the 0.15–0.3 keV luminosity, while the most conservative physically plausible estimate, using a WD atmosphere model, gives ∼4–5 (*Spectral Analysis* in *Methods*). This suggests that the most luminous HSSs found so far have bolometric luminosities of at least $10^{38}$ erg s$^{-1}$, and potentially considerably higher for cooler temperatures, given the steep temperature dependence of the correction factor in the narrow 0.15–0.3 keV band (see Extended Data Fig. 3). Even without bolometric corrections, several HSSs already approach~$10^{38}$ erg s$^{-1}$ in this band, including when accounting for lower uncertainty bounds. These results are robust to systematic effects: multiple independent approaches (see *Spectral Analysis* in *Methods*) indicate an additional ~51% systematic uncertainty, yet all plausible spectral models—not limited to thermal blackbodies—consistently predict similarly high bolometric luminosities. Intrinsic absorption that preferentially suppresses EUV emission relative to the 0.15–0.3 keV band could further reduce the bolometric correction and introduce additional systematics.

By comparison, known very soft Galactic sources such as AG Draconis (AG Dra)[12], which also emit predominantly below 0.3 keV, have much lower luminosities. The most luminous sources in NGC 4472 nearly reach ~$10^{38}$ erg $s^{-1}$ in the 0.15–0.3 keV band—almost an order of magnitude higher than AG Dra's bolometric luminosity. This discrepancy is substantial even without a bolometric correction and becomes more pronounced when accounting for the fact that any realistic spectral model—not only blackbodies— implies a non-negligible correction. For reference, AG Dra has a bolometric correction of order ~100, which raises its observed 0.15–0.3 keV luminosity from ~$2\times10^{35}$ erg $s^{-1}$ to a bolometric luminosity of ~$1\times10^{37}$ erg $s^{-1}$ (see *Methods*).

HSSs are clearly associated with the galaxies we are studying. In the elliptical galaxy NGC 3379, for instance, a substantial concentration of HSSs is observed near the galaxy's center (Fig. 1), where they account for at least 20% of the detected X-ray sources. In other elliptical or bulge-dominated galaxies, HSSs are often seen farther from the central region, yet generally remain confined within the halo. Meanwhile, in the spiral galaxy M101—the only surveyed galactic disk in our sample—HSSs primarily appear along the spiral arms (Fig. 2). The connection between the spatial distributions of HSSs and the galaxy morphologies rules out both a cosmological and a Milky Way origin, indicating that HSSs represent a distinct point-source population that urges a new classification. This conclusion is further supported by our control sample analysis of deep *Chandra* observations of non-galaxy fields (see *Control Sample* in *Methods*), which finds that foreground/background contamination is negligible (<3%). Their presence in the older stellar populations of elliptical galaxies and spiral bulges (e.g., M31) suggests that these sources are not exclusively linked to recent star formation.

Since many HSSs were only detected in merged *Chandra* observations covering many years, their time variability is difficult to assess. A few HSSs in NGC 3379 and M101 show intra-observation variability, while others remain detectable over timescales of about a year (when observed through individual exposures during that period). Intriguingly, some HSSs may completely turn off, as indicated by their absence in later observations (Fig. 3). It is also possible that some of these sources may cool to a level where they no longer emit enough 0.15–0.3 keV photons to be detected, thus only appearing to disappear. Whether HSSs are inherently persistent, transient, or recurrent remains uncertain, as their low count rates necessitate deep exposures for reliable detection. For most suitable targets—those that are nearby, large, low in hot gas content, and situated behind minimal Galactic absorption—only a single long *Chandra* observation is typically available for the entire *Chandra* mission. However, in NGC 4472, at least two HSSs exhibit clear 0.15–0.3 keV emission over a span of 11 years (Fig. 4). Although neither source showed significant emission during the relatively shorter 2010 observation, both were detected in earlier (2000; ~40 ks) and later (2011; ~136 ks) observations, suggesting possible recurrence and even persistence.

Apart from several known sources in M31 associated with post-nova systems, only a small fraction of HSSs in other galaxies have detectable counterparts at other wavelengths. In NGC 3115, for example, we identified an HSS (HSS J100514.4−074358; Extended Data Table 1) that coincides with a point source in Hubble Space Telescope (*HST*) ACS/WFC F475W imaging from March 2012 (see *Methods*). The source was undetected in X-rays before February 2012, appeared in the optical in March 2012, and was again undetected in X-rays in April 2012; there is no useful UV coverage for this object. In M101, HSS J140300.3+541952 lies in a crowded *HST* WFC3/UVIS F300X field, with at least one UV source within the $1\sigma$ X-ray positional uncertainty. Among all sources, we were able to associate only 1 of the 18 HSSs in NGC 3115 with globular clusters, in strong contrast to normal X-ray binaries (XRBs), which have long been known to be over-represented in such star clusters (~10-70%) due to likely dynamical effects within the star clusters that are conducive to forming mass-transfer binary systems[13–15]. Interestingly, an HSS in NGC 3379 (marked in yellow in Fig. 3) seems to have a UV counterpart ($0.7''$offset) associated with a young star cluster from the 2009/10 *HST* observation[16]. The 0.15–0.3 keV light curve (yellow markers) reveals that the source was X-ray bright only during the 2006 epoch and disappeared in the most recent 2007 *Chandra* observation.

Overall, HSSs are generally not associated with galactic star clusters, or any identifiable optical counterparts. However, it is important to note that most normal non–globular-cluster X-ray sources in external galaxies also lack detectable optical counterparts, so the absence of optical detections for most HSSs is not unexpected. In particular, a WD with a low-mass companion would not be detectable at optical wavelengths at distances of ~10 Mpc. Moreover, for sources hotter than ≈18 eV, a blackbody model predicts that the far-UV tail of the spectrum would fall below the detection limits of the existing *HST* data.

The physical nature (or natures) of HSSs remains unclear and are likely a heterogeneous mix. The term 'hypersoft' is purely phenomenological and may reflect an observational trait shared by multiple types of objects, rather than indicating a single, unified class. Unrelated sources may exhibit a 'hypersoft state' under different physical conditions. Many low-luminosity ($L_X \approx 10^{35}$-$10^{36}$ erg s$^{-1}$) HSSs in M31 are known post-nova systems (at least seven) and were historically classified as SSSs, but during *Chandra* observations they emit almost exclusively below 0.3 keV. For example, HSS J004255.2+412046 (kT ≈ $17^{+7}_{-6}$ eV)[11] is coincident (angular offset ~0.25″) with nova M31N 1996–08b[10], which exhibits an unusually long soft X-ray phase, turning on ~1831 days and fading ~13.8 years after the optical outburst[11].

Such systems appear as SSSs if the material ejected by the nova has become diffuse enough to allow soft X-rays to penetrate and if the WD is hot enough to emit X-rays. This occurs during the interval when the post-nova WD ceases nuclear burning and begins to cool[11,17]. As it continues to cool, a hypersoft state may be expected for some time. Indeed, we identified several sources that we classify as HSSs exhibiting such behavior. For example, in M31, a previously known SSS (CXOGSG J004232.6+411702) was later observed in a hypersoft state during subsequent *Chandra* observations.

Even more intriguing are the highest luminosity ($L_X \approx 10^{37}$-$10^{38}$ erg s$^{-1}$) sources seen in the large elliptical galaxies. Bolometrically, some of these sources may reach luminosities of $\sim 10^{39}$ erg s$^{-1}$ or higher (therefore, super-Eddington for 1M$_\odot$ object), as much total energy as ultraluminous X-ray sources (ULXs)[18,19], but exist at significantly larger numbers per galaxy (at least 20) compared to ULXs (fewer than two per galaxy, depending on star formation rate)[20] and are not confined primarily to star-forming galaxies like ULXs. In galaxies lacking an active galactic nucleus (AGN) and star formation-induced ULXs, HSSs represent the most luminous point sources in those galaxies, and yet they remained undiscovered until now because of their extremely soft X-ray emission being missed by previous surveys.

It has been conjectured that WDs that accrete and burn matter undergo thermonuclear explosions as Type Ia supernovae (SNe Ia) when they achieve the Chandrasekhar mass[21,22]. The discovery of SSSs, which have the high luminosities expected from nuclear burning, and effective radii comparable to WDs, suggested that they could be the long-sought progenitors[7]. Studies of external galaxies[23,24] concluded that SSSs are not numerous enough by about 1–2 orders of magnitude to account for the observed SNe Ia rate in galaxies. Additional work on those SSSs that appear to be steady-nuclear burners showed that some exhibit X-ray off states consistent with the expansion of the photosphere and a shift of the emission to longer, EUV wavelengths[25]. This model predicts the existence of bright HSSs, and indeed, at least one such shift has been observed from the SSS to the HSS regime in one source within NGC 4472. It may therefore be the case that nuclear-burning WDs manifest as both SSSs and HSSs.

Table 1 shows that adding HSSs increases the number of potential WD accretors by a factor of ~2 over previous estimates. Note in addition, that the detectable numbers of HSSs almost certainly represent only the tip of the population's iceberg since the EUV radiation they emit is readily absorbed. Similarly, detected SSSs constitute only a small fraction of their true populations, as the detectability of both populations is limited by observational constraints and the integrated effects of gas and dust along the line of sight[7,26]. We therefore consider the possibility that these extremely luminous HSSs could represent a significant fraction of the missing progenitors of SNe Ia. Together with the true SSS population, they may account for the total number needed to match the observed SNe Ia rate in galaxies. We note, however, that not all accreting WDs reaching the Chandrasekhar limit will necessarily produce SNe Ia, as only carbon–oxygen (CO) WDs can release sufficient nuclear energy to trigger such explosions. Since CO WDs generally form with lower masses than oxygen–neon (ONe) WDs, their growth to the Chandrasekhar mass in binary systems presents additional challenges, leaving the connection between nuclear-burning WDs and SNe Ia progenitors an open question. Therefore, arguments against soft X-ray sources being the progenitors of supernovae need to be revisited given the discovery of HSSs.

Interestingly, if the highest-luminosity HSSs ($L_X \approx 10^{37}$-$10^{38}$ erg s$^{-1}$) have blackbody temperatures $\lesssim 15$ eV, the bolometric correction for a blackbody spectrum can increase the inferred luminosity by several hundred times or more, yielding bolometric luminosities approaching $\sim 10^{40}$ erg s$^{-1}$. Even under the most conservative bolometric corrections, a luminosity much above $10^{38}$ erg s$^{-1}$ is difficult to reconcile with accreting WD models, as no WD has ever been observed to radiate at such extreme levels. This suggests the need for an alternative explanation, with an accreting BH as one possible candidate.

Regardless of their exact nature, the temperature ranges of HSSs suggest they could significantly contribute to ionizing large portions of the interstellar medium within galaxies. For decades, the detection of high-ionization emission lines, such as HeII (ionization energy of ~54 eV) in non-AGN star-forming galaxies has remained a puzzling mystery. Conventional photoionization models involving main-sequence and giant stellar populations fail to account for the HeII emission observed in both local and high-redshift galaxies. A recent study[27] proposed that soft X-ray sources such as SSSs might supply sufficient hard ionizing photons to produce these highly ionized lines. In some cases, thermal components between 5 and 20 eV were found necessary to match the observed HeII to the Hβ emission line ratios, indicating that HSSs could be instrumental in interpreting the ionization signatures detected by SDSS and *JWST* in galaxies across both low[28] and high redshifts[29]. The discovery of HSSs raises the intriguing possibility that this enigmatic class of sources may hold the key to two of astrophysics' long-standing puzzles—the origin of high-ionization lines in galaxies and the progenitors of SNe Ia.

**Table 1 | A summary of HSS/SSSs detected in six nearby galaxies.**

| Galaxy | Type | No. of Obs. | Total Exp. (ks) | D (Mpc) | $N_{\rm H}$ ($10^{20}$cm$^{-2}$) | HSSs | SSSs | $L_{\rm X}$ (HSS) (erg s$^{-1}$) |
|---|---|---|---|---|---|---|---|---|
| M31 (bulge) | Sb | 13 | 369 | 0.8 | 6.68 | 18 | 17 | $10^{35}$-$10^{36}$ |
| M101 | SABc (face-on) | 25 | 1057 | 6.5 | 1.15 | 7 | 53 | $10^{36}$-$10^{37}$ |
| NGC 3115 | E-S0 | 11 | 1139 | 9.7 | 4.61 | 18 | 12 | $10^{36}$-$10^{37}$ |
| NGC 3379 | E | 5 | 337 | 10.6 | 2.79 | 12 | 18 | $10^{36}$-$10^{37}$ |
| NGC 4697 | E | 5 | 193 | 12.5 | 2.14 | 8 | 11 | $10^{36}$-$10^{37}$ |
| NGC 4472 | E | 8 | 434 | 16.7 | 1.62 | 21 | 21 | $10^{37}$-$10^{38}$ |

**Notes:** (i) Morphological classification from the HyperLeda catalog[42]. (ii) The distances (D) are from[43-48]. (iii) $N_{\rm H}$ are estimated from published values of Galactic HI column densities[49]. (iv) HSSs are defined as sources with no significant detection above 0.3 keV and a 0.15–0.3 keV to 0.3–1.0 keV photon ratio > 8, whereas SSSs are defined as sources detected only at ≤ 1.0 keV with no emission at higher energies. (v) All $L_{\rm X}$ values correspond to the 0.15–0.3 keV energy range.

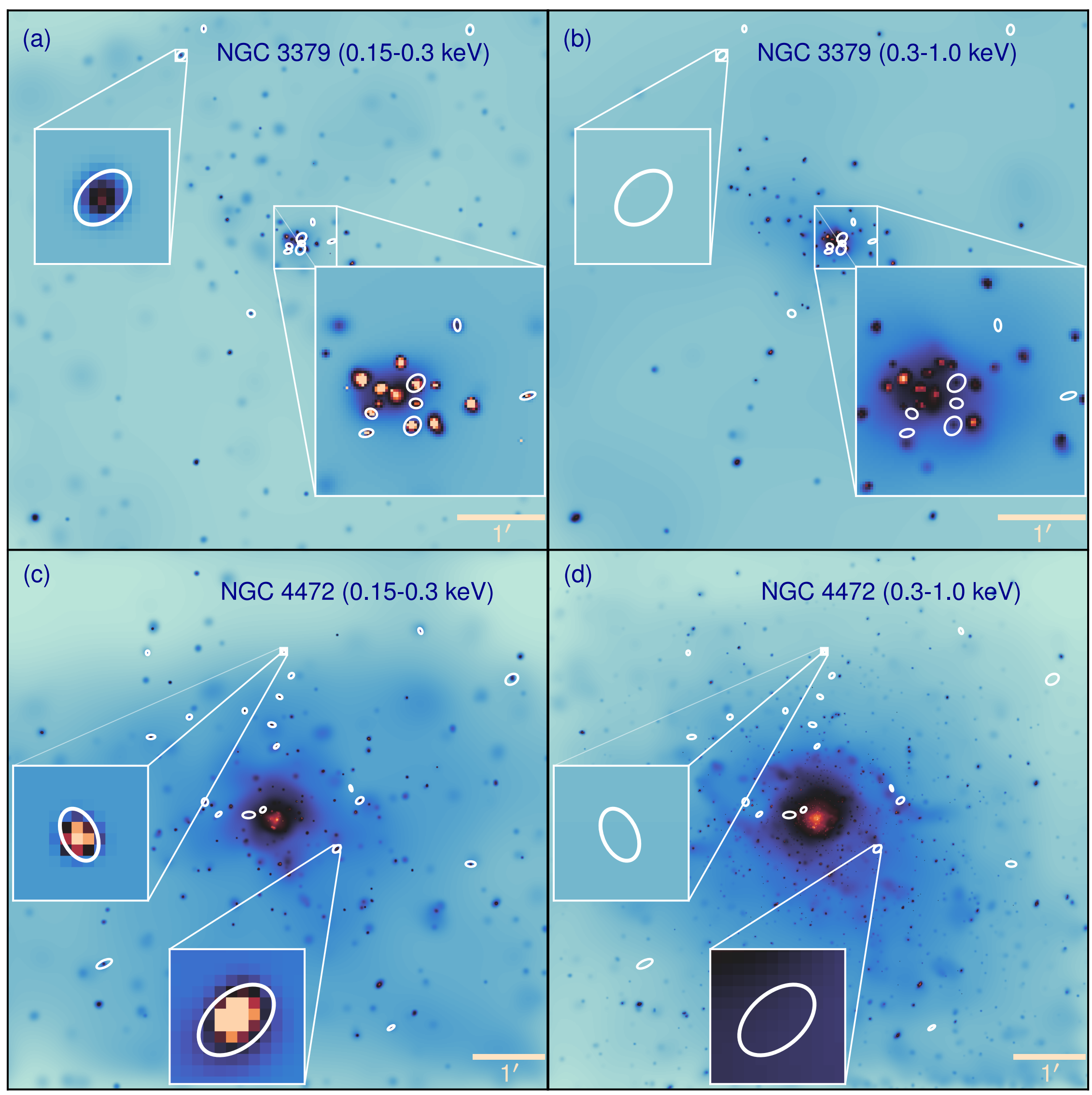

**Figure 1 | Smoothed composite *Chandra* images of NGC 3379 and NGC 4472.** (a-b) 0.15–0.3 keV and 0.3–1.0 keV band images of NGC 3379, respectively, combining 337 ks of observations. (c-d) 0.15–0.3 keV and 0.3–1.0 keV band images of NGC 4472, respectively, combining 434 ks of observations**.** The white ellipses, slightly enlarged for visual clarity, represent detections from the *Chandra* source detection tool `wavdetect`, which accounts for the point-spread function's size and orientation at each source location. They highlight the locations of the HSSs detected with at least 3σ significance in the 0.15–0.3 keV band but not above 0.3 keV. The insets show selected detections of HSSs that do not appear in the 0.3–1.0 keV bands. None of the HSSs appear in the *Chandra* Source Catalog version 2.1[50] or the comprehensive source list of[15,51], as most investigators tend to ignore energy channels below 0.3 keV.

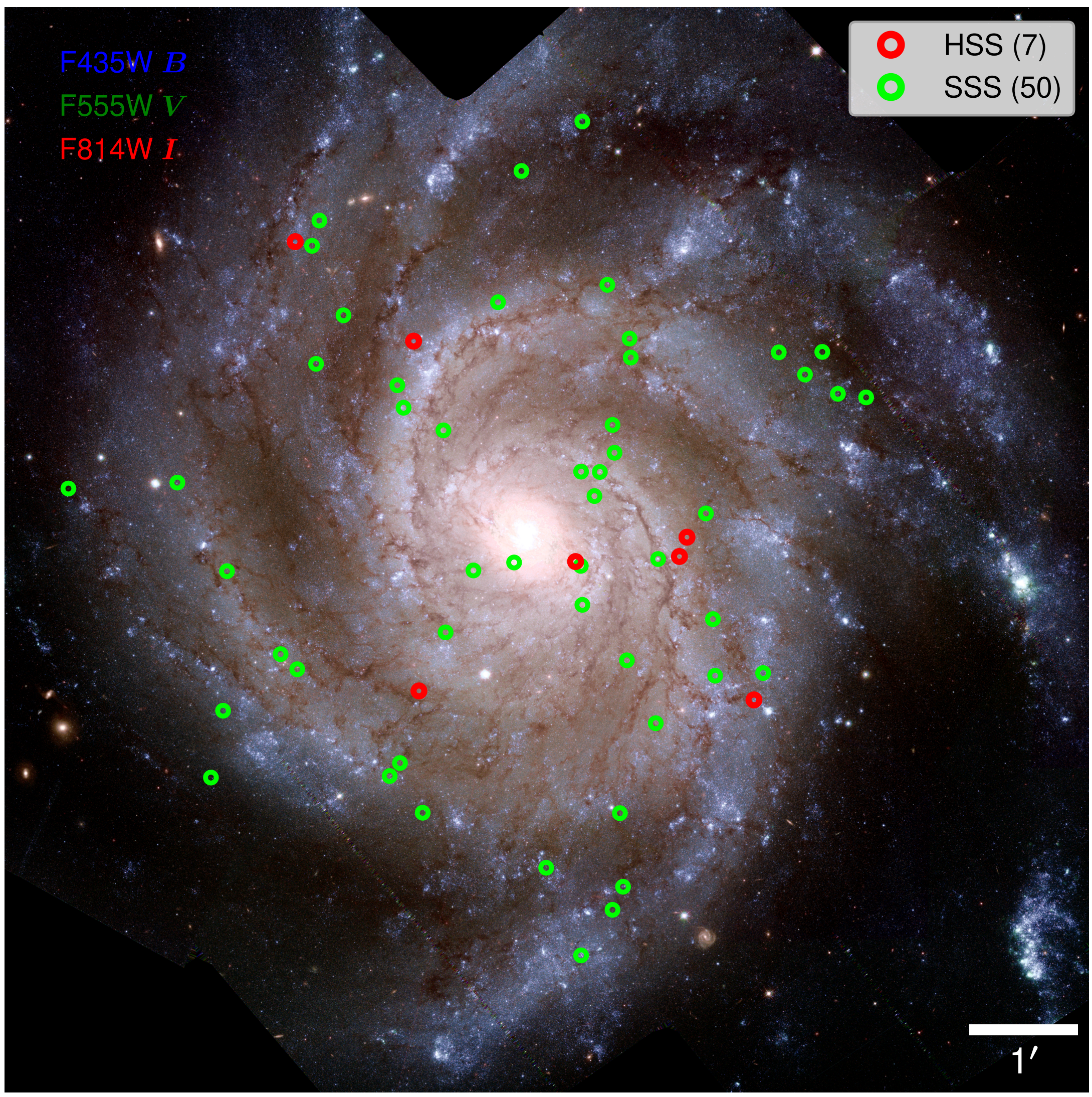

**Figure 2 | A 10′ × 10′ cutout from the full BVI color mosaic of the *HST* ACS and WFC3/UVIS observations of M101**. The location of point-like *Chandra* HSSs and SSSs are overlaid as 3″ -radius circles. The figure highlights that both types of X-ray sources primarily appear along the spiral arms of M101. The total number of sources in each category is indicated in parentheses.

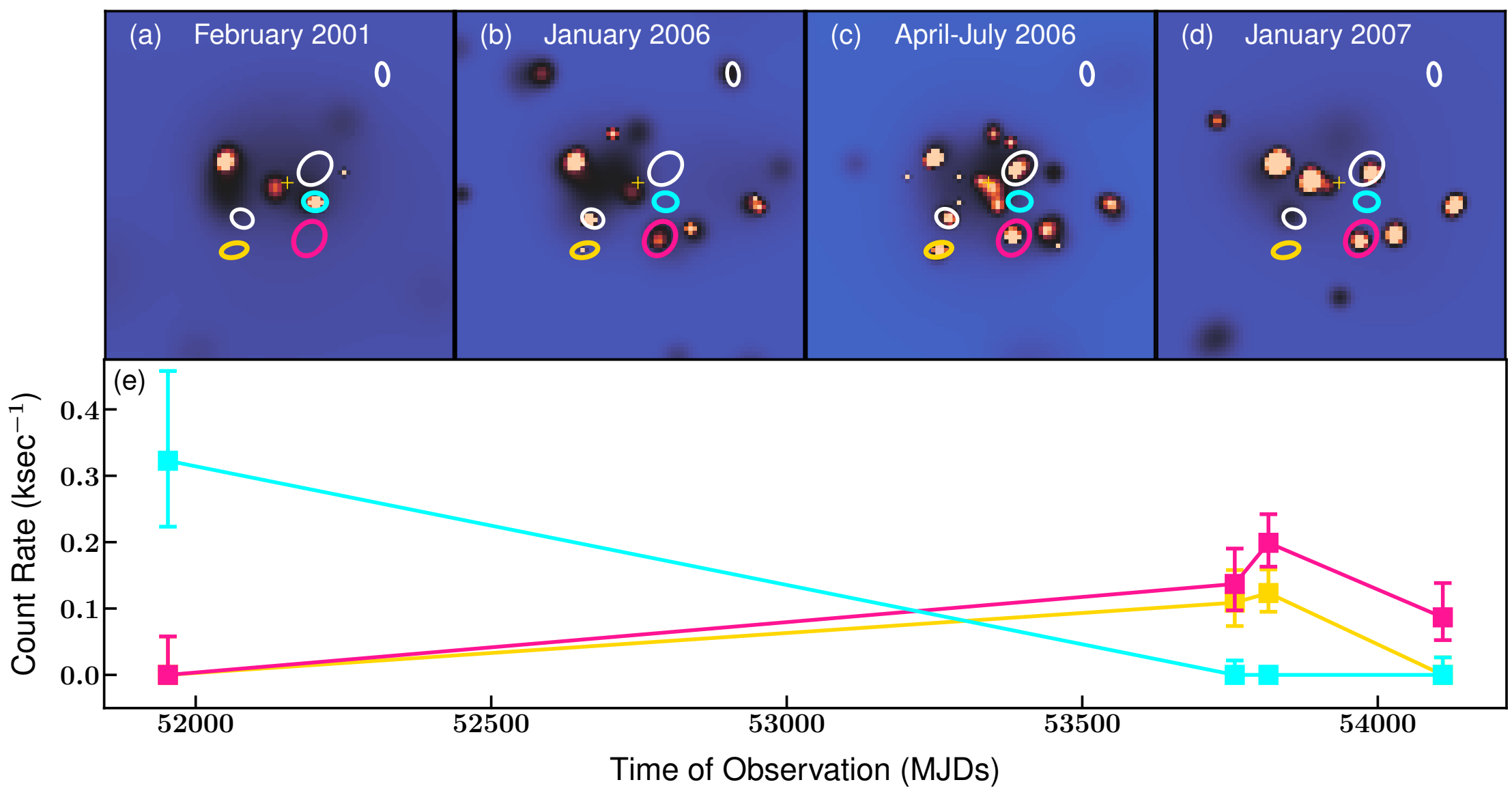

**Figure 3 | Time variability analysis of HSSs in NGC 3379.** (a-d) Adaptively smoothed 0.15–0.3 keV images of the central $40'' \times 40''$ region obtained at four epochs. Ellipses mark HSSs (as in Fig. 1), and the "+" symbol denotes the galaxy center. (c) combines two observations from 2006 (total exposure 152 ks) that were merged to improve the S/N ratio, as the HSSs remained relatively steady during this interval. One central HSS (cyan) was detected only in 2001, whereas most HSSs, such as the magenta source, were detected during multiple 2006–2007 epochs but not in 2001. The 2001 observation had a shorter exposure (32 ks). A third variable source (yellow) is spatially coincident with a UV-bright star cluster identified in 2009/10 *HST* observations[16]. (e) Sensitivity-corrected ACIS-S light curves of the three highlighted HSSs (see *Methods* for sensitivity correction details). The plots reveal that the HSSs are observed in different observations, separated by one year time scales.

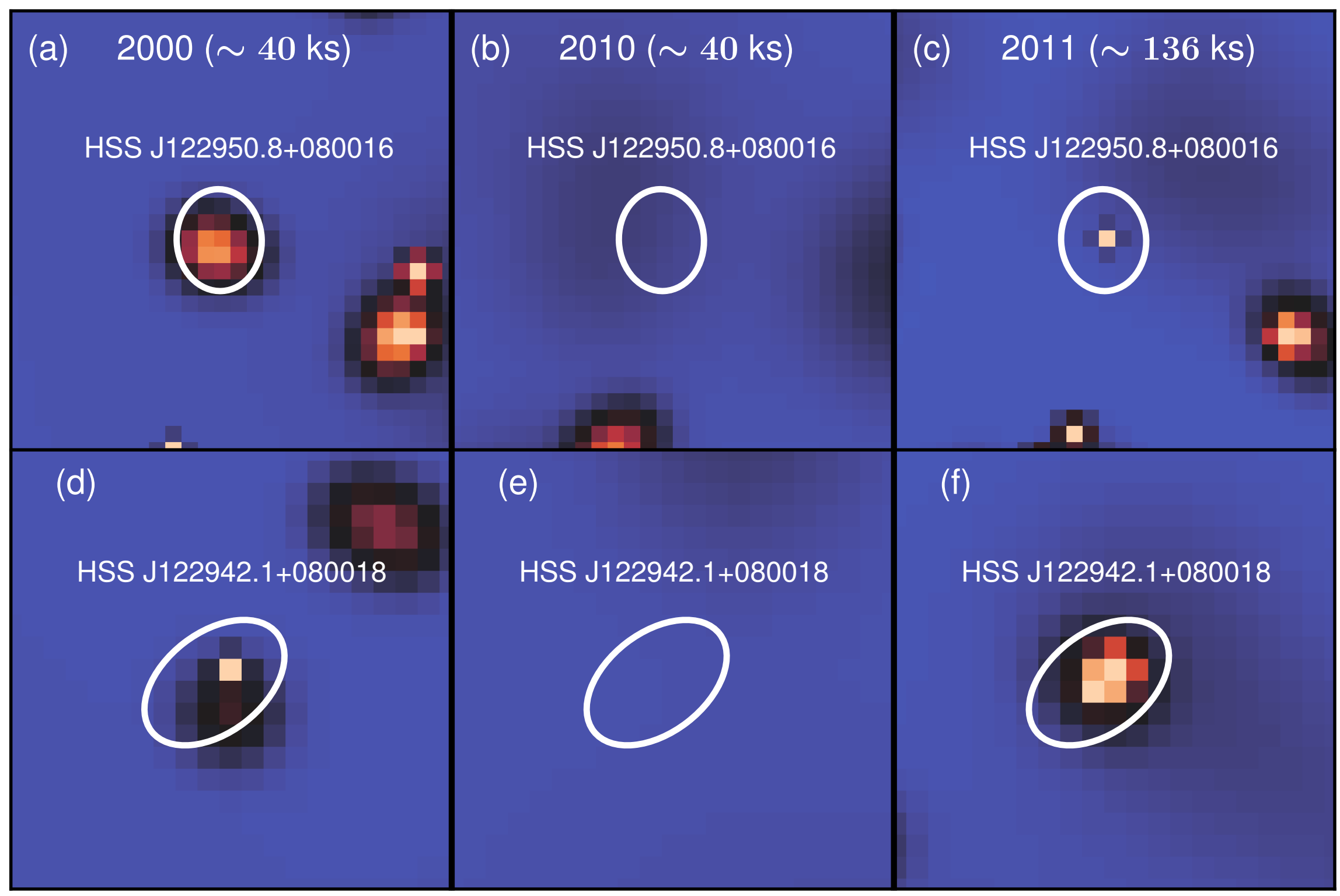

**Figure 4 | 0.15–0.3 keV *Chandra* 10″ × 10″ images of NGC 4472 across three different epochs, highlighting two HSSs that display recurrent or potentially persistent behavior.** The ellipses represent the PSF size and orientation at each source location from `wavdetect`. (a–c) HSS J122950.8+080016, located northeast of the galactic center, and (d–f) HSS J122942.1+080018, positioned to the northwest. Both are visible in 2000 (~40 ks) and 2011 (~136 ks) but absent in 2010. This non-detection in 2010 was attributed to the degradation of *Chandra*'s ACIS-S soft X-ray sensitivity, which necessitates longer exposure times in later years.

# Methods

## *Chandra* Observation and Data Reduction

Due to the accumulation of contaminants on *Chandra*'s Advanced CCD Imaging Spectrometer (ACIS) detectors over the years, which absorb soft photon energies, the detection of HSSs through the 0.15–0.3 keV channel has become increasingly difficult in recent years. Therefore, we used archival ACIS-S images and event files of the galaxies listed in Table 1, up to the year 2017 (with justification for this limit provided later). These data were reprocessed and calibrated using `CIAO` (version 4.16) and `CALDB` (version 4.11.55), following standard data reduction procedures from the *Chandra* X-ray Center (CXC). This process ensures the correction of charge-transfer inefficiencies and bad pixels and creates a new flare-cleaned level-2 event file for each observation. For each galaxy, we subsequently merged available observations using the script *merge_obs*, which reprojects and combines observations to the same tangent point, creating merged event files and images. The point-spread function (PSF) size for a given image was determined based on the 80% enclosed counts fraction at each pixel location.

## Computing ACIS-S's Time-Dependent Sensitivity

To estimate the changes in ACIS-S's sensitivity since its launch, we consider the galaxy cluster Abell 1795, a non-variable cosmic X-ray source, as our master calibrator. We found 18 ACIS-S observations with a total exposure of 238 ks in the archive where the cluster centroid ($\alpha_{J2000.0} = 13^h 48^m 52.70^s$, $\delta_{J2000.0} = +26^\circ 35' 27''$) does not fall off the chip. These observations were reprocessed using a procedure similar to the one previously discussed. For each event file, we created 0.15–0.3 keV and 0.3–1.0 keV threshold images and exposure maps, using effective energies of 0.2 keV and 0.6 keV, respectively. These threshold images were used to calculate photon counts within a $20''$ extraction circle centered on the cluster. Net photon counts were derived after subtracting the background estimated from an annulus (180–220″), where the X-ray surface brightness is <3% of that in the central 20″ and is dominated by residual cluster soft excess, particle, and cosmic X-ray backgrounds. We also calculated the average effective exposure times from the exposure maps by extracting average values from the same circles and normalizing them by the maps' maximum values. Dividing net photon counts by the average effective exposure gives exposure-normalized counts, which we later re-normalized to the first observation time.

In Extended Data Fig. 1, we show these re-normalized counts as a function of observatory age (in modified Julian dates). The figure clearly demonstrates a significant reduction in re-normalized counts in later years, particularly after Cycle 18 (year 2017), where the values drop below 20% of the original counts in Cycle 1 (year 1999) in both the 0.15-0.3 keV (red) and 0.3-1.0 keV bands (lime). Therefore, we set Cycle 18 as our limiting cycle, as discussed earlier, because beyond this point, the detectability of HSS photons becomes extremely challenging.

After considering several models to fit these data, we found that quartic polynomial functions best represent the sensitivity degradation for both bands, with reduced chi-square statistics of $\chi_\nu^2 \approx 0.34$ and $\chi_\nu^2 \approx 1.2$, respectively. These models are shown as dashed red and dotted lime curves in the figure. We later utilized these models in our analysis to account for the degradation of the sensitivity of ACIS-S over time.

**HSS Identification**

To identify HSS candidates in each galaxy, we initially run `wavdetect`[30] on the merged 0.15–0.3 keV images, following[31]. We used wavelet scales of 1, 2, 4, 8, and 16 pixels. A point source is considered as an HSS candidate when it shows counts below 0.3 keV energy, with a minimum of a 3σ detection above the local background. Additionally, we require no significant detection (< 3σ) above 0.3 keV. This evaluation is performed over a background area at least five times the size of the region enclosing 80% of the source's PSF. In general, these background extraction regions are circular annuli concentric with the source regions. However, in crowded areas, the background regions were modified to ensure that they did not overlap with any source regions. In contrast, SSS candidates are identified if they are detected in ≤ 1.0 keV but lack any detection signals in higher energy bands.

We performed the same procedure on individual observations to increase our sample completeness, ensuring that we did not miss any source that was detectable in a specific epoch but undetectable in merged observations due to added background noise from additional observations. As an additional classification criterion, we require the 0.15–0.3 keV to 0.3–1.0 keV photon ratio to exceed 8—whether measured in individual or merged observations—for a source to be classified as an HSS (see *Spectral Analysis* for justification). Sources with ratios below this threshold are instead classified as SSSs if they lack detectable emission above 1 keV.

To improve sample purity, each candidate was visually inspected to remove potential false positives arising from hot pixels, extended galactic structures, or poorly sampled background regions, reducing the initial list from 125 to 84 sources. Because *Chandra* ACIS-S's nominal reliable detection threshold is 0.2 keV—slightly above our chosen 0.15 keV threshold—we verified whether these 84 sources remain detectable in the 0.2–0.3 keV band. We found that 81 of 84 sources are still detected, indicating that the difference in effective collecting area between the two bands affects only <4% of sources. Consequently, we adopt these 84 sources as our final HSS sample. Source positions were subsequently shifted to the *Gaia*[32] reference frame via astrometric correction, as listed in Extended Data Table 1. With at least 3σ detections above their local background, HSSs are unlikely to be detector artifacts or cosmic-ray afterglows. Multiple-epoch detections further rule out Poisson fluctuations coinciding at the same location. Moreover, they cannot be clumps of low S/N hot gas in the galaxies since they are found more readily in nearby hot gas-poor galaxies.

For the brighter HSSs, spectral variations were investigated where the data permitted; consequently, a small fraction (< 10%) of HSSs may exhibit phase transitions. In subsequent observations, the source appeared to return to its HSS state, indicating a possible phase transition from HSS to SSS and back to HSS. Furthermore, some HSSs in our sample from M31 were previously classified as low-temperature SSSs (with optical nova counterparts) in earlier X-ray studies[11,17,33]. These classifications were based on detections in the 0.2–1.0 keV bands from *XMM-Newton*, *Chandra*, and ROSAT observations, although most of their emissions were concentrated in the 0.2–0.3 keV band.

**Control Sample**

To assess whether a significant fraction of HSSs could arise from foreground or background sources, we performed a survey sensitivity analysis incorporating a control sample. Field sensitivity depends on several factors, including exposure time, field of view, and, most critically, the Galactic column density ($N_H$). As illustrated in Extended Data Fig. 1, exposure times must be corrected for the long-term degradation of the *Chandra* ACIS-S detector; we therefore adopt an effective exposure time that accounts for this effect using the fitted profile in Extended Data Fig. 1.

For the control sample, we selected 16 non-galaxy fields with long *Chandra* observations spanning 1999–2017, matching the epochs of our galaxy fields and reaching comparable sensitivity. While we detect 84 HSSs across six galaxies, applying the same selection criteria to the control fields yields only two HSS candidates, both with low count rates. This indicates that the vast majority of HSSs in our galaxy sample originate from the galaxies themselves, consistent with the spatial distributions described in the main text. Based on the control sample, we estimate that fewer than ~3% of HSSs (2 of 84) could be attributed to foreground or background sources such as AGN or Galactic objects.

### Counterpart Identification

In NGC 3115, we identified an *HST* counterpart to the highly variable HSS J100514.4−074358 in an ACS/WFC F475W image from March 2012, after registering both datasets to the *Gaia* frame, with a residual relative offset of ≈0.31″. The source was not detected in X-rays prior to February 2012, appeared in the optical in March 2012, and was again undetected in X-rays in April 2012. In M101, HSS J140300.3+541952 lies in a crowded *HST* WFC3/UVIS F300X field, with at least one UV source within the 1σ X-ray positional uncertainty after registering both images to the *Gaia* frame (offset ≈0.29″). In NGC 3379, HSS J104749.9+123447 appears to have a UV counterpart in *HST* WFC3/UVIS F225W imaging, consistent with a source reported earlier[16]. However, tying the *HST* image to the *Gaia* frame is challenging due to the paucity of UV sources in the field, preventing a robust astrometric verification. Additional candidate counterparts may be present in M101, NGC 4472, and NGC 4697, and will be investigated in a future study.

### Spectral Analysis

Fitting a spectrum and deriving the X-ray luminosity for an HSS is challenging, given the limited calibration data available for the ACIS detectors below 0.3 keV. As a more robust way to obtain fluxes of our HSS candidates, we derived a counts-to-energy flux conversion factor in the 0.15–0.3 keV range using normal XRBs that emit in both the 0.15–0.3 keV and harder (e.g., 0.5–6.0 keV) bands. For each target galaxy, we extracted the full spectra from these XRBs using `specextract` and combined them into a single spectrum to boost photon statistics in the 0.15–0.3 keV band, thereby enhancing the statistical reliability of the analysis. We then fit an absorbed power-law model in the 0.5–6.0 keV range using XSPEC[34], obtaining a best-fit photon index in the range 1.5–1.8 for six galaxies. The model was extrapolated to estimate the unabsorbed flux in the 0.15–0.3 keV band, correcting for variations of $N_H$ toward each target galaxy. By dividing this unabsorbed flux by the total photon counts in the 0.15–0.3 keV range from the combined spectrum and multiplying it by the total exposure time, we derived a 0.15–0.3 keV counts-to-energy flux conversion factor. This factor was then applied to the count rates of HSSs identified in each galaxy to estimate their unabsorbed luminosities in the 0.15–0.3 keV band.

To assess the robustness of the conversion factor, we fitted an absorbed thermal *APEC* model to the hot gas in NGC 4472 over the 0.5–6.0 keV range, including an additional power-law component to account for residual XRB emission. The resulting 0.15–0.3 keV count-to-energy flux conversion factor differed by no more than 25% from that obtained using the power-law model fitted to the XRB population.

As a third independent calibration test, we fit absorbed blackbody models to SSSs in M31. Unlike many Galactic SSSs, these sources are largely unaffected by pile-up and have sufficient counts above 0.3 keV for direct spectral fitting. Extrapolating the best-fit models to 0.15 keV yielded direct estimates of the 0.15–0.3 keV flux and an independent count-to-energy flux conversion factor. After accounting for differences in exposure time, assumed column density ($N_H$), and contamination buildup, the conversion factors derived from the three methods differed by no more than 51%, indicating that the conversion factor is modestly dependent on the assumed spectral model over the narrow 0.15–0.3 keV band. We therefore adopt a 51% systematic uncertainty in the conversion factor and propagate it to the unabsorbed flux and luminosity estimates. This is consistent with previous calibration studies showing that low-energy calibration effects are most significant for high-count spectra with small statistical uncertainties[35], whereas the sources considered here are photon-limited.

Following the methods discussed above, we derive the unabsorbed 0.15–0.3 keV luminosities in the range of $10^{35}$– $10^{36}$ erg $s^{-1}$ for M31 HSSs, assuming spherically symmetric emission, the host-galaxy distance, and column densities from the literature, with no intrinsic source absorption. In contrast, large elliptical galaxies such as NGC 4472 and NGC 3379 harbor HSSs with luminosities in the range of $10^{36}$– $10^{37}$ erg $s^{-1}$, with some even approaching $10^{38}$ erg $s^{-1}$. These high luminosities within such a narrow energy band imply potentially higher bolometric luminosities, depending on the underlying source properties and temperatures. However, very low count rates in this band, together with calibration uncertainties of ACIS-S below 0.3 keV, make it difficult to constrain the temperatures, particularly if the emission peaks in the EUV.

Rather than attempting to derive temperatures of HSSs, our goal was to constrain an upper limit on their effective temperature by comparing the 0.15–0.3 keV to 0.3–1.0 keV photon ratios of HSSs with those of SSSs, for which temperatures are better constrained owing to detectable emission above 0.3 keV that enables spectral fitting. For example, ultraviolet spectroscopy performed with the Cosmic Origins Spectrograph (COS) aboard the *HST* led to demonstrate that RX J0439.8-6809—a persistent bright SSS identified by ROSAT and located in the Milky Way galactic halo—is likely associated with a hot Milky Way WD with a temperature of 21.5 ± 2.6 eV (250000 ± 30000K)[9]. The interstellar hydrogen column density ($N_H$) toward this source is estimated to be $5 \times 10^{20} \mathrm{cm}^{-2}$ [9], which falls within the range of Galactic column density estimates for the HSSs in our sample (Table 1). An early *Chandra* ACIS-S observation revealed that this source has a 0.15–0.3 keV to 0.3–1.0 keV photon ratio of 5.7:1.

Summing the 0.15–0.3 keV and 0.3–1.0 keV counts for all the HSSs in NGC4697, NGC3379, and NGC3115 (the cleanest, minimally absorbed galaxies in our sample) yielded $570 \pm 32$ and $69 \pm 27$ counts, respectively, for the two bands. This yielded a 0.15–0.3 keV to 0.3–1.0 keV ratio of $8.2 \pm 3.2$:1. Note that, although individual HSSs may not exhibit significant 0.3–1.0 keV emission, the cumulative signal from multiple sources can achieve statistical significance. We therefore adopt a photon ratio threshold of ≈8 as an additional criterion for HSS classification. The data indicate that HSSs exhibit a mean 0.15–0.3 keV to 0.3–1.0 keV photon ratio slightly higher than that of RX J0439.8−6809, implying mean temperatures below its upper-limit value of 24 eV; otherwise, detectable Wien-tail emission would be present above 0.3 keV. Therefore, we estimate that HSSs are no hotter than 24 eV based on the lack of their 0.3–1.0 keV emission, with some potentially being considerably cooler ($\sim$15 eV).

Notably, in such a narrow band, where the bulk of the emission peaks in the EUV, the spectral peaks of cooler sources shift toward longer wavelengths (lower energies), as illustrated by the idealized blackbody flux function in Extended Data Fig. 2. As a result, the bolometric correction becomes strongly dependent on the assumed equivalent temperature (Extended Data Fig. 3). For instance, the bolometric corrections to the 0.15–0.3 keV luminosities for 25 eV, 24 eV, 20 eV, 15 eV, and 10 eV blackbodies are 7, 8, 18, 105, and 4794, respectively (derived from XSPEC blackbody (BB) models; cyan curve in Extended Data Fig. 3b), assuming all fluxes are unabsorbed ($N_H$ corrected).

Among the known M31 HSSs associated with recent novae, only one source has a published temperature estimate: M31N 1996–08b, with a blackbody temperature of $17^{+7}_{-6}$ eV based on *XMM-Newton* observations[10,11,36], consistent with the upper limit of RX J0439.8−6809. As shown in Extended Data Fig. 3, the corresponding bolometric correction factors span ~8–1600 across the allowed temperature range when considering both the upper and lower uncertainties, assuming a blackbody spectrum. In contrast, the SSSs from M31 (according to our classification scheme) have higher temperature estimates, with the lowest being $30^{+4}_{-4}$ eV (M31N 2000-07a)[11], corresponding to a much smaller blackbody bolometric correction factor of ~3–6. On the rarest occasions, if the sources are associated with a hot WD such as RX J0439.8−6809, a WD atmosphere model predicts a bolometric correction of ~4–5 at 24 eV, which serves as a conservative lower bound. Although blackbody fits generally yield systematically lower temperatures and higher luminosities than atmospheric models at these temperatures, the blackbody approximation is adopted here for simplicity and primarily to facilitate comparison among sources. We emphasize that the intrinsic SEDs of the HSSs remain unknown; therefore, all bolometric corrections discussed here are inherently model dependent and should be regarded as illustrative rather than definitive.

Consequently, the highly luminous sources in galaxies such as NGC 3379 and NGC 4472 could have bolometric luminosities exceeding $10^{38}$ erg s$^{-1}$ and possibly approaching $10^{39}$ erg s$^{-1}$, depending on their exact temperatures. This suggests that the physical origin of these highly luminous, low-temperature HSSs could be fundamentally different from that of M31 HSSs, which are primarily associated with novae. Interestingly, if an HSS has a blackbody temperature below 18 eV, the Rayleigh-Jeans tail of its spectrum could potentially be detected in the FUV (900–2150 Å or 0.0058–0.0138 keV) using the *HST*.

To place the luminosities of the brightest NGC 4472 HSSs in context, we compare them with the well-studied very soft Galactic source AG Dra[12]. A study[37] reports that AG Dra reaches a ROSAT PSPC count rate of ~1 ct s$^{-1}$ at peak. Using PIMMS and assuming a 15 eV blackbody with $N_H$ = 3.1 × $10^{20}$ cm$^{-2}$, this corresponds to a 0.15–0.3 keV flux of 8.2 × $10^{-11}$ erg cm$^{-2}$ s$^{-1}$, or $L_X$ ~ 2.2 × $10^{35}$ erg s$^{-1}$ at a distance of 4.8 kpc. Applying a bolometric correction of ~100 for a 15 eV blackbody yields a bolometric luminosity of ~$10^{37}$ erg s$^{-1}$. By comparison, the most luminous NGC 4472 HSSs already emit near $10^{38}$ erg s$^{-1}$ in the narrow 0.15–0.3 keV band. Adopting a nominal equivalent temperature of 24 eV implies a minimum bolometric correction factor of ~8 for a blackbody—the smallest value expected for any reasonable spectral model— while multicolor disk (MCD)[38] models indicate substantially larger corrections, typically 3–4 times higher (Extended Data Fig. 3b). This shows that many NGC 4472 HSSs are likely more luminous than AG Dra even without bolometric corrections.

**Estimating Uncertainties**

To account for the low counts in HSSs, $1\sigma$ uncertainties in the measured counts were estimated using a Bayesian approach appropriate for small-number Poisson statistics[39]. In short, for each source, we computed the posterior probability distribution of the true source counts given the observed total counts in the source and background regions, adopting a non-informative prior for the source intensity and accounting for the relative areas and exposures of the two regions[40]. Credible intervals were then derived directly from the posterior. These were then propagated to 0.15–0.3 keV X-ray fluxes, including the previously noted 51% systematic uncertainties in the counts-to-flux conversion. Luminosity uncertainties further incorporate the galaxy distance uncertainties listed in Supplementary Data Table 1 but do not account for uncertainties in the assumed column density.

**Data Availability:** The *Chandra* data supporting the findings of this study are available in the *Chandra* Data Archive under the identifier doi:10.25574/cdc.395.

**Code Availability:** The code used to generate catalogs, figures, and relevant results in this study are additionally accessible from Zenodo: doi: 10.5281/zenodo.20534943 [41].

**Correspondence:** Correspondence and requests for materials should be addressed to Mustafa Muhibullah (mmuhibullah@crimson.ua.edu).

**Acknowledgments:** This scientific work uses data obtained from archival *Chandra* X-ray Center observations, which is operated by the Smithsonian Astrophysical Observatory for and on behalf of NASA under contract NAS8-03060. This work was supported by the *Chandra* X-ray Center under grant AR4-25004X (M.M., J.A.I.). R.D.S. would like to acknowledge support from NSF AST-2009520.

**Author Contributions:** J.A.I., M.M., and R.D.S. jointly conceived the project. M.M. and J.A.I. developed the methodology. The investigation was carried out by M.M., J.A.I., and R.D.S. Data visualization was performed by M.M., J.A.I., and R.D.S. M.M. and J.A.I. secured *Chandra* funding for the project. Project administration was handled by J.A.I. and M.M., with supervision provided by J.A.I., R.D.S., and M.M. The original manuscript draft was written by M.M., R.D.S., and J.A.I., and all authors contributed to reviewing and editing the final version.

**Competing Interests:** The authors declare no competing interests.

# Extended Data

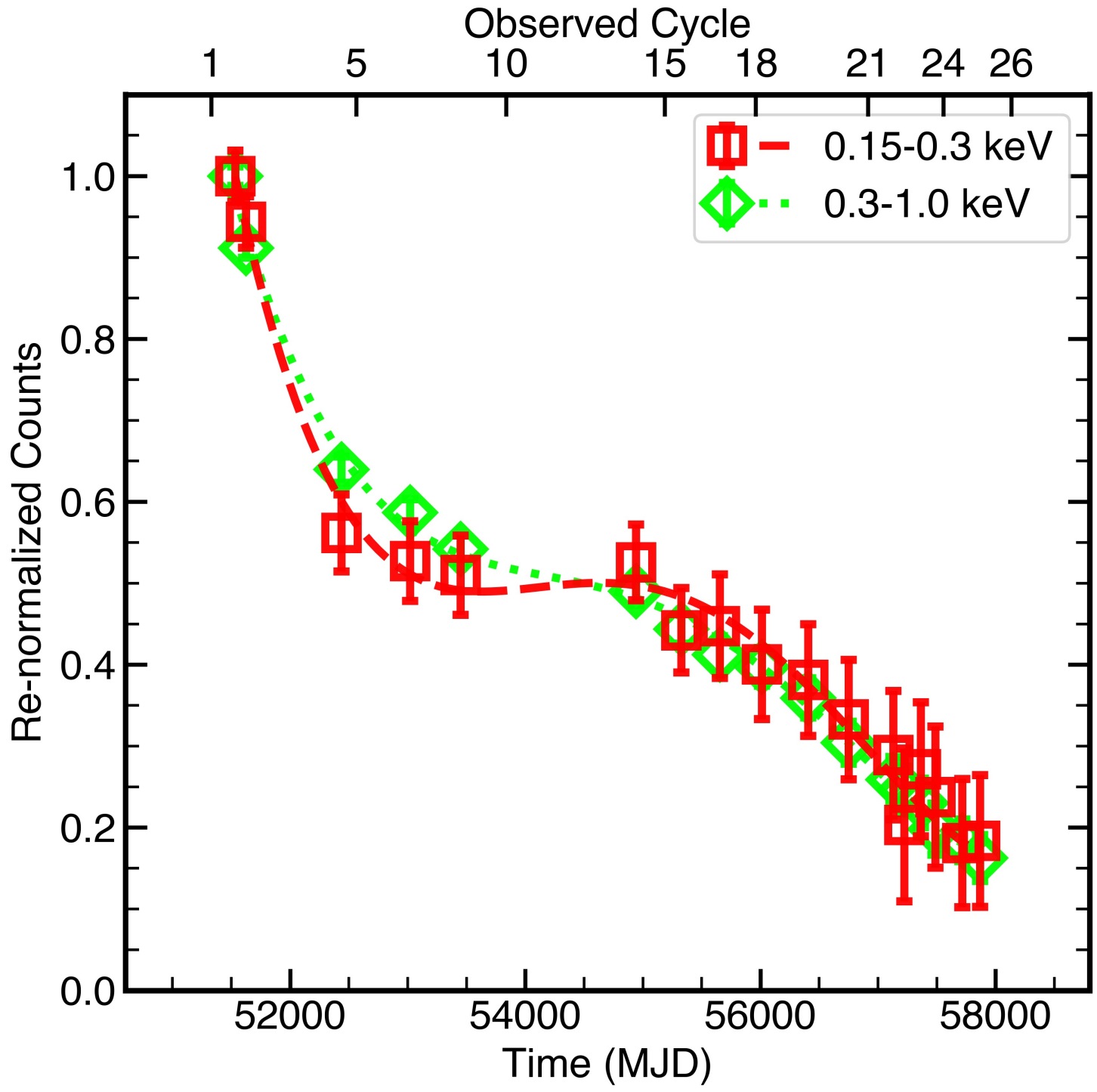

**Extended Data Figure 1: Exposure re-normalized number counts in the 0.15–0.3 keV and 0.3–1.0 keV bands, estimated from Abell 1795 as a function of Modified Julian Date (MJD) and *Chandra* observing cycle.** The plot illustrates the progressive decline in *Chandra*'s sensitivity over time, particularly in the softest energy channels. The error bars represent $\mathbf{1\sigma}$ uncertainties, with central values referring mean values. The dashed red curve and dotted lime curve correspond to the best-fit quartic polynomial models with $\chi^2_\nu \approx \mathbf{0.34}$ and $\chi^2_\nu \approx \mathbf{1.2}$, respectively. Interestingly, the ratio of 0.15–0.3 keV to 0.3–1.0 keV counts remains relatively consistent over time.

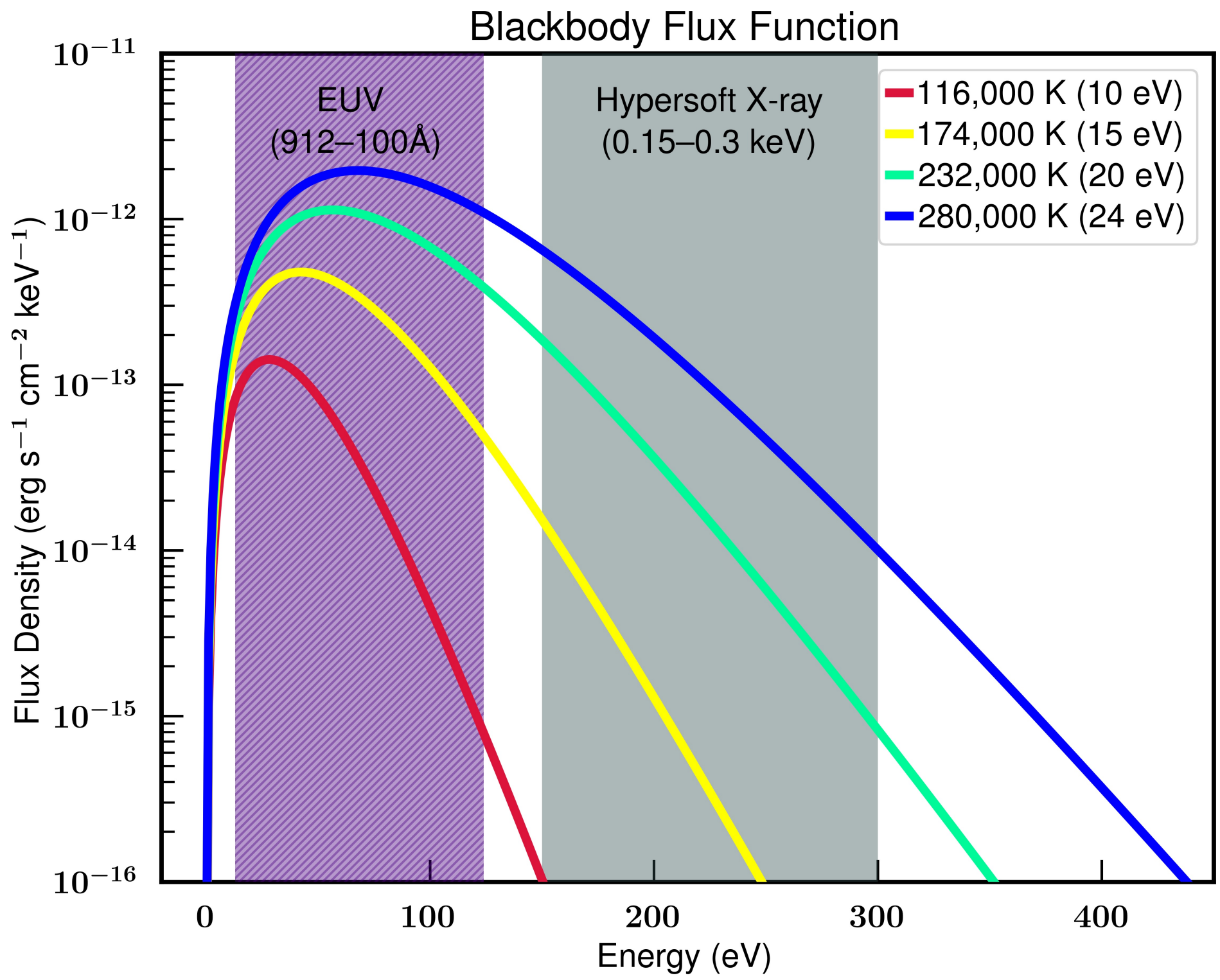

**Extended Data Figure 2: Ideal blackbody flux density as a function of photon energy for sources with different blackbody temperatures.** Each curve represents a thermal emitter within the expected temperature range of the HSSs, illustrating the shift of the spectral peak toward lower energies for cooler sources. The shaded regions mark the EUV (912–100 Å) and 0.15–0.3 keV bands. The flux densities are scaled to span the range of observed HSS fluxes in this study, highlighting the fraction of emission falling within the 0.15–0.3 keV band.

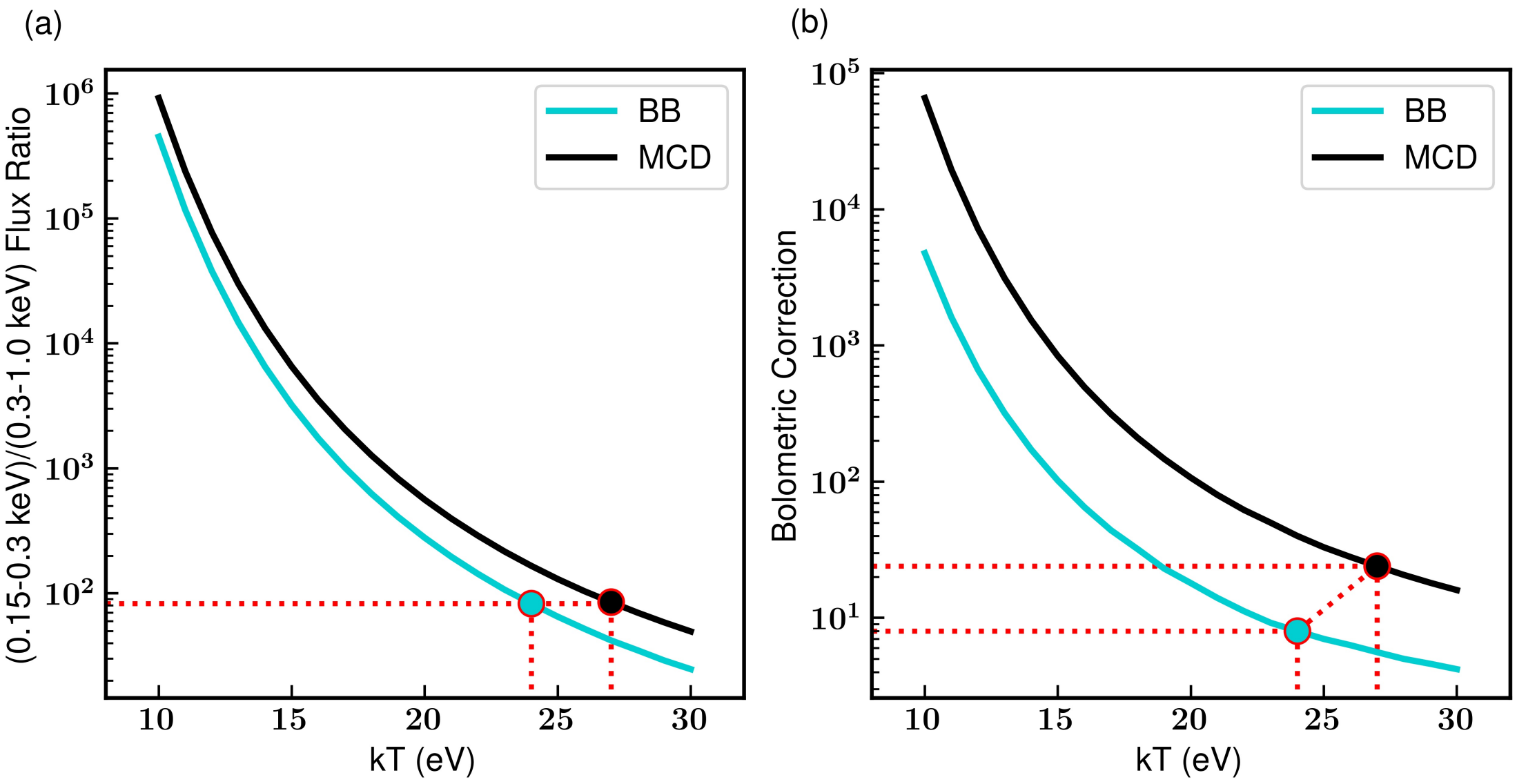

**Extended Data Figure 3: Model-predicted flux ratios and bolometric corrections vs. source temperature.** (a) The 0.15–0.3 keV to 0.3–1.0 keV flux ratio for the blackbody (BB) and multicolor disk (MCD) models, respectively. The horizontal dashed red line indicates the flux ratio at which HSSs are expected to be hottest ($\mathbf{kT_{BB} \approx 24}$ eV), corresponding to a slightly higher temperature for the accretion disk model ($\mathbf{kT_{MCD} \approx 27}$ eV), (b) shows that the bolometric correction for the MCD model is approximately 3–4 times higher than that of the BB model at the respective temperatures.

**Extended Data Table 1 | Properties of HSSs discovered in this work.**

A portion of the full catalogue is shown here for guidance regarding its form and content. The complete table is provided as Supplementary Data Table 1, and a machine-readable version is available at https://zenodo.org/records/20534943/files/ed_table1.fits?download=1.

| Galaxy | Distance (Mpc) | Object | Good Exp. (ks) | RA (J2000) | Dec (J2000) | Net Counts (0.15-0.3kev) | Net Err. | Flux (0.15-0.3kev) (erg/s/cm$^2$) | Flux Err. (erg/s/cm$^2$) | $L_X$ (0.15-0.3keV) (erg/s) | $L_X$ Err. (erg/s) | Counterpart Candidates ($r \leq 0.5''$) | SSS Phase |
|---|---|---|---|---|---|---|---|---|---|---|---|---|---|
| M31 | 0.8 | HSS J004245.0+411521 | 369.5 | 00:42:45.01 | +41:15:21.02 | 112.69 | 11.0, -10.26 | 4.52E-14 | 1.21E-14, -1.20E-14 | 3.04E36 | 8.52E35, -8.45E35 | Nova 2009-05a (Henze+,2014), XMM/EPIC-pn, Chandra/HRC-I | ObsIDs: 13826 (2.6$\boldsymbol{\sigma}$), 14197 (2.4$\boldsymbol{\sigma}$) |
| M101 | 6.5 | HSS J140300.3+541952 | 917.2 | 14:03:00.35 | +54:19:51.58 | 34.98 | 6.73, -6.08 | 5.26E-15 | 1.66E-15, -1.60E-15 | 2.66E37 | 8.67E36, -8.38E36 | UV match (HST/WFC/UVIS) | - |
| NGC3115 | 9.7 | HSS J100514.4-074358 | 1138.7 | 10:05:14.44 | −07:43:58.41 | 20.71 | 5.04, -4.41 | 2.33E-15 | 8.15E-16, -7.67E-16 | 2.63E37 | 1.06E37, -1.01E37 | Optical match (HST/ACS/WFC) | - |
| NGC3379 | 10.6 | HSS J104749.3+123456 | 337.1 | 10:47:49.31 | +12:34:56.01 | 38.34 | 6.69, -6.08 | 4.42E-15 | 1.35E-15, -1.31E-15 | 5.95E37 | 2.17E37, -2.13E37 | SSS (Wang+, 2016) | ObsID: 7074 (1.95$\boldsymbol{\sigma}$) |
| NGC4697 | 11.4 | HSS J124833.5-054816 | 193.0 | 12:48:33.52 | −05:48:15.68 | 11.55 | 3.85, -3.18 | 1.28E-15 | 5.31E-16, -4.75E-16 | 2.39E37 | 1.01E37, -9.01E36 | - | - |
| NGC4472 | 16.7 | HSS J122943.4+075937 | 428.7 | 12:29:43.39 | +07:59:37.44 | 23.69 | 5.44, -4.89 | 3.99E-15 | 1.35E-15, -1.29E-15 | 1.33E38 | 4.61E37, -4.42E37 | - | - |

**Notes:** (i) All coordinates are shifted to the *Gaia* reference frame[32]. (ii) Asymmetric (upper and lower) 1σ uncertainties in net counts (Net Err.) are estimated from the Bayesian posterior probability distribution[39,40]. (iii) The unabsorbed estimated 0.15–0.3 keV flux. (iv) The unabsorbed 0.15–0.3 keV X-ray luminosities ($L_X$) assume spherical symmetry, the host-galaxy distance, and column densities from the literature, with no intrinsic source absorption; $<3\%$ of sources are expected to be foreground/background based on control analysis. (v) Flux and luminosity uncertainties include a 51% systematic uncertainty in the flux conversion factor. Luminosity uncertainties also include host-galaxy distance uncertainties from the literature (negligible), but exclude uncertainties in the assumed column density. (vi) Counterpart candidates denote sources matched within $r \leq 0.5''$ of the X-ray position, including literature and newly identified counterparts from *XMM*/EPIC-pn, *Chandra*/HRC-I, and *HST* observations used in this study. (vii) “SSS Phase” indicates observations in which the HSS exhibited a supersoft spectral state, including the detection significance.